\RequirePackage{ifpdf}
\documentclass[12pt]{JHEP3}         
\usepackage{amssymb,amsfonts}
\usepackage{cite}
\usepackage{amsmath}
\usepackage{graphicx}
\usepackage{MnSymbol}
\usepackage[utf8]{inputenc}

\def\rmd{{\rm d}}

\newcommand{\ke}{\rangle}
\newcommand{\br}{\langle}

\newcommand{\ba}{\begin{eqnarray}}
\newcommand{\ea}{\end{eqnarray}}
\newcommand{\be}{\begin{equation}}
\newcommand{\ee}{\end{equation}}
\newcommand{\bal}{\begin{align}}
\newcommand{\eal}{\end{align}}
\newcommand{\bay}[1]{\left(\begin{array}{#1}}
\newcommand{\eay}{\end{array}\right)}

\newcommand{\st}[1]{|#1\ke}

\newcommand{\Tr}{\mbox{Tr}}

\def\xD{{\Delta}}

\def\CN{{\cal N}}

\def\IH{\Bbb{H}}
\preprint{TAUP-3001/15}
\title{R\'enyi entropy of free $(2,0)$ tensor multiplet and its supersymmetric counterpart}

\author{Jun Nian${}^{1,2}$, Yang Zhou${}^{3}$ \\

${}^1$Institut des Hautes \'Edutes Scientifiques\\ Le Bois-Marie, 35 route de Chartres, 91440 Bures-sur-Yvette, France\\

${}^2$C.N. Yang Institute for Theoretical Physics\\ Stony Brook University, Stony Brook, NY 11794-3840, U.S.A.\\

${}^3$School of Physics and Astronomy, Tel-Aviv University\\ Ramat-Aviv 69978, Israel\\

{\tt E-mails: nian@ihes.fr, yangzhou@post.tau.ac.il}
}

\abstract{We compute the R\'enyi entropy and the supersymmetric R\'enyi entropy for the six-dimensional free $(2,0)$ tensor multiplet. We make various checks on our results, and they are consistent with the previous results about the $(2,0)$ tensor multiplet. As a by-product, we have established a canonical way to compute the R\'enyi entropy for $p$-form fields in $d$-dimensions. 
}

\begin{document}

\pagestyle{plain} \setcounter{page}{1}
\newcounter{bean}
\baselineskip16pt


\section{Introduction}
Entanglement entropy (EE) and R\'enyi entropy have been intensively studied in recent years. They not only play important roles in quantum information theory and condensed matter physics, but also bring new insights into high energy physics. For instance, in the context of conformal field theories they are related to the conformal anomaly \cite{CHM} and can be computed holographically \cite{RT}. Combining the ideas of supersymmetric localization \cite{Witten, Nekrasov, Pestun} and R\'enyi entropy, one can also define the supersymmetric refinement of the ordinary R\'enyi entropy on the branched sphere \cite{YN}. Interestingly, the supersymmetric R\'enyi entropy enjoys universal relations with central charges in even dimensions, which provides a new way to derive the Hofman-Maldacena bounds~\cite{Zhou:2015cpa}.

Let us briefly review these concepts. Suppose the space on which the theory is defined can be divided into a piece $A$ and its complement $\bar{A}=B$, and correspondingly the Hilbert space factorizes into a tensor product $\mathcal{H} = \mathcal{H}_A \otimes \mathcal{H}_B$. The density matrix over the whole Hilbert space is $\rho$; 
then the reduced density matrix is
\be
  \rho_A := \textrm{tr}_B \rho\, .
\ee
The entanglement entropy is the von Neumann entropy of $\rho_A$,
\be
  S_E := - \textrm{tr} \rho_A \, \textrm{log} \, \rho_A\, ,
\ee
while the R\'enyi entropies are defined to be 
\be
  S_n := \frac{1}{1-n} \textrm{log}\, \textrm{tr} (\rho_A)^n\, .
\ee
Assuming that a satisfactory analytic continuation of $S_n$ can be obtained, we can alternatively express the entanglement entropy as the $n\to 1$ limit of the R\'enyi entropies:
\be
  \lim_{n\to 1} S_n = S_E\, .
  \label{analyticcont}
\ee

In the flat space $\Bbb{R}^{1,d-1}$, the R\'enyi entropy of a $d$-dimensional conformal field theory $S_n$ across a spherical entangling surface $\Bbb{S}^{d-2}$ can be mapped to that on a branched sphere, $\Bbb{S}^d_n$, with the same entangling surface on the north pole. Throughout this work we will take the ``regularized cone'' boundary condition~\cite{Lewkowycz:2013laa}.
The R\'enyi entropy can also be computed using the thermal partition function on a hyperbolic space $\Bbb{S}^1_n\times\Bbb{H}^{d-1}$, where the entangling surface is mapped to the boundary of $\Bbb{H}^{d-1}$. The universal part of $S_n$ is shown to be invariant under Weyl transformations of the metric~\cite{CHM}:
\ba
S_n &=& {1\over 1-n}\left(\log Z_n[\Bbb{S}^d_n] - n\log Z_1[\Bbb{S}^d]\right) \label{renyi1}\\ &=& {1\over 1-n}\left(\log Z_n[\Bbb{S}^1_n\times\IH^{d-1}] - n\log Z_1[\Bbb{S}^1\times\IH^{d-1}]\right)\ .\label{renyi2}
\ea These identities are generally true for both non-supersymmetric and supersymmetric R\'enyi entropies of all CFTs.

The concept supersymmetric R\'enyi entropy was first studied in three dimensions~\cite{YN, Huang:2014gca}, and later generalized to four dimensions~\cite{Huang:2014pda,Zhou:2015cpa,Crossley:2014oea} and five dimensions~\cite{Alday:2014fsa,Hama:2014iea}.
By turning on certain background gauge fields (chemical potentials) and using the supersymmetric localization technique, one can calculate the partition function $Z_n$ on the $n$-branched sphere, and define the supersymmetric R\'enyi entropy:
\be\label{sre}
  S_n^{SUSY} = \frac{\textrm{log} Z_n (\mu(n)) - n\, \textrm{log} Z_1 (0)}{1-n}\, .
\ee
It is a supersymmetric refinement of the ordinary R\'enyi entropy, which is in general non-supersymmetric because of the conic singularity. The quantities defined by (\ref{sre}) are in general UV divergent, but one can extract universal parts free of ambiguities. This becomes particularly clear in even dimensions. For instance, for $\CN=4$ SYM in four dimensions, the log coefficient of supersymmetric R\'enyi entropy as a function of $n$ and three chemical potentials $\mu_1, \mu_2, \mu_3$ (corresponding to three $U(1)$ Cartans of $SO(6)$ R-symmetry respectively) has been shown to be protected from the interactions~\cite{Huang:2014pda}. It also received a precise check from the holographic computation of the BPS 3-charge topological AdS black hole in five-dimensional gauged supergravity~\cite{Huang:2014pda}. The above facts indicate that supersymmetric R\'enyi entropy may be used as a new robust observable to understand superconformal field theories (SCFTs).

Our main concern in this work is six-dimensional $(2,0)$ superconformal field theories. While it is easy to identify a free Abelian tensor model that realizes $(2,0)$ superconformal symmetry, the existence of interacting $(2,0)$ theories was only inferred from their embedding into particular constructions in string theory~\cite{Witten:1995zh,Strominger:1995ac,Witten:1995em}. It is not yet understood how to formulate interacting $(2,0)$ SCFTs. Furthermore, it is believed that the relevant or marginal deformations preserving $(2,0)$ supersymmetry do not exist~\cite{Cordova:2015vwa,Intriligator15talk}. Therefore, as the initial start we will mainly focus on free Abelian tensor multiplets in the present work. 

There are some previous works on the R\'enyi entropy of $6d$ CFTs. In~\cite{Myers} and~\cite{Safdi:2012sn}, the shape dependence of entanglement entropy for general $6d$ CFTs was initially studied~\footnote{See~\cite{Miao:2015iba} for further investigations.} and some concrete results of entanglement entropy for $(2,0)$ theories were also presented in~\cite{Safdi:2012sn}. Based on these works, in~\cite{Lewkowycz:2014jia} the authors investigated the shape dependence of R\'enyi entropy, where they pointed out that the $n$-dependence of R\'enyi entropy across a non-spherical entangling surface (with vanishing extrinsic curvature) is actually determined by that across a spherical one in flat space.

In this note we provide the first direct field-theoretic calculation of the R\'enyi entropy and the supersymmetric R\'enyi entropy for the most interesting six-dimensional SCFTs, the $(2, 0)$ theories. We compute the R\'enyi entropy $S_n$ explicitly by carefully analyzing the contribution of the two-form field with self-dual strength. The result receives quite a few consistency checks. Namely, our R\'enyi entropy result at $n=1$ is consistent with~\cite{Bastianelli:2000hi,Safdi:2012sn}, while the first derivative $\partial_n S_n$ at $n=1$ and the two-point correlator of stress tensor~\cite{Bastianelli:1999ab} satisfy the relation proposed in~\cite{Perlmutter:2013gua}. Furthermore, the second derivative $\partial_n^2 S_n$ at $n=1$ and the three-point correlator of stress tensor~\cite{Bastianelli:1999ab} satisfy the relation proposed in~\cite{Lee:2014zaa}. We also obtain the supersymmetric counterparts of the R\'enyi entropy for the tensor multiplet, which behave reasonably simple as a function of $n$ because of supersymmetry.

The note is organized as follows. In Section 2, we calculate the R\'enyi entropy for the six-dimensional $(2, 0)$ tensor multiplet, and perform various checks on the result.
In Section 3, we turn on the chemical potential and compute the supersymmetric R\'enyi entropy for several cases. We conclude in Section 4 and leave the explicit analysis of Killing spinor equations in Appendix \ref{KSEa}.

\section{R\'enyi entropy of 6$d$ free CFTs}
\label{renyifreefields}
The R\'enyi entropy of 6$d$ CFTs in $\Bbb{R}^6$ associated with a 4$d$ spherical entangling surface $\Bbb{S}^4$ can be computed using the thermal partition function on a hyperbolic space $\Bbb{S}^1\times\Bbb{H}^5$. The partition function $Z(\beta)$ on $\Bbb{S}^1_\beta\times\Bbb{H}^5$ can be computed from the heat kernel of the gapless  Laplacian $\Delta$,
\be\label{partitionkernel}
\log Z(\beta) = {1\over 2} \int_0^\infty {dt\over t}K_{\mathbb{S}^1_\beta\times \mathbb{H}^5}(t)\ ,
\ee where $K(t)$ is defined as the trace of the kernel of the operator $\Delta$
\be
K(t):= \Tr (e^{- t \xD}) = \int \rmd^6 x\sqrt g\, K(x,x,t)\ ,\quad
K(x,y,t) := \br x|e^{- t \xD}\st{y}\ .
\ee Since $\Bbb{S}^1_\beta\times\Bbb{H}^5$ is a direct product, the kernel is factorized,
\be
K_{\mathbb{S}^1_\beta\times \mathbb{H}^5}(t)=K_{\mathbb{S}^1_\beta}(t)\, K_{\mathbb{H}^5}(t)\ .
\ee
The heat kernel on $\mathbb{S}^1$ is given by
\be\label{S1kernel}
K_{\mathbb{S}^1_\beta}(t)={\beta\over \sqrt{4\pi t}}\sum_{m\neq 0,\in\mathbb{Z}} e^{-\beta^2m^2\over 4t} \ .
\ee The hyperbolic space $\mathbb{H}^5$ is homogeneous and therefore the volume $V_5$ factorizes
\be
K_{\mathbb{H}^5}(t)= \int d^5x \sqrt{g}~K_{\mathbb{H}^5}(x,x,t):=V_5\, K_{\mathbb{H}^5}(0,t)\ ,
\ee where $V_5=\pi^2\log(\ell/\epsilon)$ is the regularized volume of $\mathbb{H}^5$. $\epsilon$ is the UV cutoff of the theory and $\ell$ is the curvature radius of the hyperbolic space.

\subsection{Complex scalar}
The heat kernel of a complex scalar on $\Bbb{H}^5$ is given by
\be
K^s_{\Bbb{H}^5}(t)={V_5\over (4\pi t)^{5/2}} \left(2+{4t\over 3}\right)\ .
\ee The free energy can be computed by (\ref{partitionkernel})
\be
F^s(\beta):=-\log Z(\beta) = {\pi V_5\over 1890\beta^5}(-8\pi^2-7\beta^2)\ ,
\ee and the R\'enyi entropy is given by
\be\label{renyis}
S^s_n :={n F(\beta=2\pi) - F(\beta=2n\pi)\over 1-n} = \frac{(n+1) \left(3 n^2+1\right) \left(3 n^2+2\right) V_5}{15120 \pi ^2 n^5}\ .
\ee This reproduces the R\'enyi entropy result of conformal scalars first presented in~\cite{Casini:2010kt}.
\subsection{Weyl fermion}
The heat kernel of a Weyl fermion on $\Bbb{H}^5$ is given by (we have taken into account $ 2^{[{6\over 2}]}/2=4$ components)
\be
K^f_{\Bbb{H}^5}(t)={V_5\over (4\pi t)^{5/2}} \left(4+{20t\over 3}+3 t^2\right)\ .
\ee The free energy can be computed by (\ref{partitionkernel}) with anti-periodic boundary conditions along $\Bbb{S}^1$
\be
F^f(\beta)= {-V_5\over 60480\pi\beta^5}(496\pi^4+980\pi^2\beta^2+945\beta^4)\ ,
\ee and the R\'enyi entropy is given by
\be\label{renyif}
S^f_n = \frac{(n+1) \left(1221 n^4+276 n^2+31\right) V_5}{120960 \pi ^2 n^5}\ .
\ee 
This reproduces the R\'enyi entropy of massless fermions first presented in~\cite{Lee:2014zaa}.
\subsection{Two-form}
One has to be careful about the heat kernel computation of R\'enyi entropy for $p$-form fields. To do this, we employ the general results of eigenvalue distributions for Hodge-de Rham operator of $p$-form fields in $N$-dimensional hyperbolic space~\cite{CamporesiHiguchi},
\be\label{pformkernel}
\Tr\, K_{\mu_1\cdots\mu_p}^{~~~~~\nu_1\cdots\nu_p}(x,x^\prime,t) = V_N\int_0^\infty d\lambda\,\,\mu(\lambda)\,e^{-[\lambda^2+(\rho-p)^2]\,t}\ ,
\ee
where $\rho:={N-1\over 2}$ and the trace has been taken for both indices $\mu_1\cdots\mu_p$ and coordinates $x$. The eigenvalue distribution $\mu(\lambda)$ is given by
\be
\mu(\lambda) = {c_N\over \Omega_{N-1}}{\pi\, g(p)\over \left[2^{N-2}\Gamma({N\over 2})\right]^2\left[\lambda^2+(\rho-p)^2\right]} \prod_{j=0}^{N-1\over 2}(\lambda^2+j^2)\qquad (N~\text{odd})\ ,
\ee where $c_N$, $g(p)$ and $\Omega_{N}$ are defined as
\be
c_N := {2^{N}\over 4\pi}\ , \quad g(p) := {(N-1)!\over p!(N-p-1)!}\ ,\quad \Omega_N := {2\pi^{N+1\over 2}\over \Gamma({N+1\over 2})}\ .
\ee

One may want to first reproduce the known R\'enyi entropy result of gauge field in 4$d$ by using (\ref{pformkernel}). In that case, $p=1$ and $N=3$. Therefore,
\be
c_3={2\over \pi}\ ,\quad g(p=1)=2\ ,\quad \Omega_2=4\pi\ .
\ee
The eigenvalue distribution $\mu(\lambda)$ is
\be
\mu(\lambda) = {\lambda^2+1\over \pi^2}\ .
\ee
The heat kernel of a 1-form field on $\Bbb{H}^3$ is given by
\be
K^v_{\Bbb{H}^3}(t) = V_3\frac{2 t+1}{4 \pi ^{3/2} t^{3/2}}\ .
\ee
With this kernel the R\'enyi entropy can be computed by evaluating the partition function (\ref{partitionkernel})
\be
S_{d=4}^v = \frac{(n+1) \left(31 n^2+1\right) V_3}{360 \pi  n^3}\ .
\ee After adding up with a constant discrepancy between entanglement entropy $S_{n\to 1}$ and 4$a_4$, where $a_4$ is the standard gauge field trace anomaly coefficient, we are able to reproduce the known R\'enyi entropy of a 4$d$ gauge field
\be
S^v_{d=4} = \frac{\left(91 n^3 + 31 n^2+n+1\right) V_3}{360 \pi  n^3}\ .
\ee
Notice that the physical degrees of freedom of the 4$d$ Abelian gauge field (photon) is two, hence we learn from the 4$d$ exercise that the formula (\ref{pformkernel}) actually takes into account all the physical degrees of freedom for the $p$-form field.

Now we turn to two-form field in 6$d$, which corresponds to $p=2$ and $N=5$. In this case,
\be
c_5={8\over \pi}\ ,\quad g(p=2)=6\ ,\quad \Omega_4= {8\pi^2\over 3}\ .
\ee
The eigenvalue distribution is
\be
\mu(\lambda) = {(\lambda^2+1)(\lambda^2+4)\over 2\pi^3}\ .
\ee
The kernel of a 2-form field on $\Bbb{H}^5$ is given by
\be
K^v_{\Bbb{H}^5}(t) = V_5 \frac{2 t (8 t+5)+3}{16 \pi ^{5/2} t^{5/2}}\ .
\ee
With this kernel the R\'enyi entropy is obtained as
\be\label{renyivtest}
S_{d=6}^v = \frac{(n+1) \left(877 n^4+37 n^2+2\right) V_5}{5040 \pi ^2 n^5}\ .
\ee
Notice that this is not the final result of the R\'enyi entropy for a 2-form field in 6d, as one can check that there is an apparent discrepancy between $S_{EE}:=S^v_{n\to1}$ and 4$a^v_6$, where $a^v_6$ is the standard 2-form field trace anomaly coefficient in 6$d$. According to the general proof of the universal relation~\cite{CHM} between entanglement entropy and trace anomaly for any CFT, one has to take a constant shift of (\ref{renyivtest}) to get the correct value of entanglement entropy (R\'enyi entropy at $n=1$). We will see that the same constant shift also gives the correct R\'enyi entropy (for $n\neq 1$) which survives quite a few consistency checks. As discussed in \cite{Huang:2014pfa}, this constant shift comes from the gauge redundancy on the boundary of the entangling surface, hence it is independent of the parameter $n$. Instead of carefully analyzing the gauge redundancy on the boundary, we will
simply take the shift by matching to the trace anomaly of the 2-form field. We will see that this recipe indeed gives the correct result.

To figure out the discrepancy between the entanglement entropy of a 2-form field and its trace anomaly coefficient 4$a_6$, we make use of the ratios between the trace anomalies of one complex scalar field, one Weyl fermion and one two-form field~\cite{Bastianelli:2000hi},
\be\label{ratio1}
a^s_6: a^f_6 : a^v_6 = {5\times 2\over 72} : {191\over 72\times 2} : {221\over 4}\ . 
\ee
The ratios between the entanglement entropies of one complex scalar, one Weyl fermion and one two-form field from our results (\ref{renyis})(\ref{renyif}) are
\be\label{ratio2}
S^s_{n\to 1}:S^f_{n\to 1}:S^v_{n\to 1} = {2\over 756} {V_5\over \pi^2} : {191\over 7560} {V_5\over \pi^2}: S^v_{n\to 1}\ .
\ee
By requiring the equality of (\ref{ratio1}) and (\ref{ratio2}), we can obtain
\be
S^v_{n\to 1} = {221\over 210}{V_5\over \pi^2}\ .
\ee
Comparing with the $n\to 1$ value of (\ref{renyivtest}), we obtain the discrepancy
\be
\Delta S^v = \left({221\over 210} - {229\over 630}\right){V_5\over \pi^2} = {434\over 630}{V_5\over \pi^2}\ .
\ee Therefore, the correct R\'enyi entropy of a 2-form field in 6d should be
\be
S^v_n = S_{d=6}^v + \Delta S^v = \frac{(n+1) \left(37 n^2+2\right)+877 n^4+4349n^5}{5040 n^5}{V_5\over \pi^2}\ .
\ee
This is one of our new results.
\subsection{(2,0) tensor multiplet}
A six-dimensional $(2,0)$ tensor multiplet includes five real scalars, two Weyl fermions and one 2-form field with self-dual strength. The 2-form field with self-dual strength can also be considered as a chiral 2-form field which has half of the degrees of freedom. Putting the contributions of all fields together, we get the R\'enyi entropy of the $(2,0)$ tensor multiplet
\be
S^{(2,0)}_n = 5\times {S^s_n\over 2} + 2 S^f_n + {S^v_n\over 2} = {(n+1)(28 n^2 + 3) + 313 n^4 + 1305 n^5\over 2880 n^5}{V_5\over \pi^2}\ .
\ee
The entanglement entropy of the $(2,0)$ tensor multiplet is given by
\be\label{eemultiplet}
S^{(2,0)}_{EE} = S^{(2,0)}_{n\to 1} = {7\over 12}{V_5\over \pi^2}\ .
\ee The first and second derivatives at $n=1$ are
\be
\partial_n S^{(2,0)}_n \big|_{n=1} = -{1\over 6}{V_5\over \pi^2}\ ,\quad \partial^2_n S^{(2,0)}_n \big|_{n=1} = {4\over 9}{V_5\over \pi^2}\ . 
\ee
A few consistency checks are in order.
\begin{itemize}
\item  The entanglement entropy (\ref{eemultiplet}) is consistent with the result (2.12) in~\cite{Safdi:2012sn} for a spherical entangling surface $\Bbb{S}^4$, if one takes into account a factor difference ${16\over 7}$N$^3$~\cite{Bastianelli:2000hi} between the trace anomaly of $(2,0)$ tensor multiplet and that of the large-N theory of coincident M5-branes. It is also consistent with (2.29) in~\cite{Bastianelli:2000hi} if one adopts the normalization condition $-{1\over 8\times 3!}{1\over (4\pi)^3} \int_{\mathbb{S}^6} E_6=2$.

\item The first derivative $-{1\over 6}{V_5\over \pi^2}$ is consistent with the coefficient of the two-point correlator of the stress tensor in the $(2,0)$ tensor multiplet, which is given by~\cite{Bastianelli:1999ab}
\be\label{CT}
C_T = {84\over \pi^6}\ .
\ee
By consistency we mean the universal relation between the first derivative of R\'enyi entropy at $n=1$ and $C_T$ in any CFT as shown in~\cite{Perlmutter:2013gua}
\be\label{SnCT}
S^\prime_{n=1} = -\text{Vol}(\Bbb{H}^{d-1}){\pi^{d/2+1}\Gamma(d/2)(d-1)\over (d+1)!} C_T\ .
\ee
Using (\ref{CT}) we indeed find that our result $\partial_n S^{(2,0)}_n \big|_{n=1} = -{1\over 6}{V_5\over \pi^2}$ satisfies the relation (\ref{SnCT}) for $d=6$.

\item The second derivative ${4\over 9}{V_5\over \pi^2}$ is consistent with the coefficients of the three-point correlator of the stress tensor in the $(2,0)$ tensor multiplet. The relation between $S''_{n=1}$ and the coefficients $\mathcal{A}$, $\mathcal{B}$ and $\mathcal{C}$ in the three-point correlator of the stress tensor was derived in \cite{Lee:2014zaa}:
\be\label{SnABC}
  S''_{n=1} = {4 \pi^{d+1}\over 3 d^3 (d+2) \, \Gamma(d-1)}\, \textrm{Vol} (\mathbb{H}^{d-1}) \left((4 d^2 - 10 d + 8) \mathcal{A} - d \mathcal{B} - (10 d - 8) \mathcal{C} \right)\, .
\ee
The coefficients $\mathcal{A}$, $\mathcal{B}$ and $\mathcal{C}$ for the $(2,0)$ tensor multiplet are given in \cite{Bastianelli:1999ab}:
\be\label{ABC}
  \mathcal{A} = - {2^6\cdot 3^4\over 5^2 \pi^9}\, ,\quad \mathcal{B} = - {181\cdot 2^4 \cdot 3^2 \over 5^2 \pi^9}\, ,\quad \mathcal{C} = - {59\cdot 2^3\cdot 3^3\over 5^2 \pi^9}\, .
\ee
As discussed in \cite{Lee:2014zaa}, for the complex scalar there is a mismatch between the known result for $S''_{n=1}$ and the one derived from the right hand side of \eqref{SnABC}. For $d=6$ this mismatch is compensated by a factor ${113\over 125}$. Since we have calculated the R\'enyi entropies for the complex scalar and for the $(2, 0)$ tensor multiplet explicitly, the expected result for (\ref{SnABC}) is given by
\be
  {113\over 125} \cdot {5\over 2} \partial^2_n S^s_n \big|_{n\to 1} + \left[\partial^2_n S^{(2,0)}_n \big|_{n\to 1} - {5\over 2} \partial^2_n S^s_n \big|_{n\to 1} \right] = {11\over 25} {V_5 \over \pi^2}\, .
\ee
One can check that this indeed agrees with the right hand side of \eqref{SnABC} by using (\ref{ABC}).
\end{itemize}

\section{Supersymmetric R\'enyi entropy}
Now we study the supersymmetric R\'enyi entropy of the free tensor multiplet. To preserve supersymmetry, one has to turn on an R-symmetry background gauge field to twist the boundary conditions for scalars and fermions along the replica direction~\cite{Belin:2013uta}. In the manifold $\Bbb{S}^1_{\beta=2\pi n}\times\Bbb{H}^5$, this is equivalent to turning on an R-symmetry chemical potential along $\Bbb{S}^1_\beta$, therefore the heat kernel along the circle has a phase shift. The kernel (\ref{S1kernel}) becomes
\be\label{twistkernel}
\widetilde K_{\mathbb{S}^1_\beta}(t)={\beta\over \sqrt{4\pi t}}\sum_{m\neq 0,\in\mathbb{Z}} e^{{-\beta^2m^2\over 4t}+i2\pi m\mu-i\pi mf} \ ,
\ee 
where $f=0$ for scalars and $f=1$ for fermions. Using this twisted kernel one can perform the same computations of partition functions as what has been done in Section \ref{renyifreefields}. These partition functions are functions of the chemical potential $\mu$. In consideration of supersymmetry, $\mu$ has to be a function of $n$ and it should be vanishing at $n=1$ since we do not need additional background field for the manifold $\Bbb{S}^1_{\beta=2\pi}\times\Bbb{H}^5$, which is conformally equivalent to a round six-sphere. The explicit function $\mu(n)$ can be found by solving Killing spinor equation either on a branched $\Bbb{S}^6_n$ or on $\Bbb{S}^1_n\times\Bbb{H}^5$ since it is invariant under Weyl transformation.

The supersymmetric R\'enyi entropy is defined as
\be
S^{SUSY}_n := {nF_1(0)-F_n(\mu)\over 1-n}\ .
\ee
Similar to the case in 4$d$, it is also convenient to extract the extra contribution due to the nontrivial chemical potential,
\be
\Delta S := S^{SUSY}_n - S_n\ .
\ee
For a complex scalar,
\be\label{extras}
\Delta S^s(\mu) = \frac{V_5\, \mu ^2 (\mu +1)^2 \left(-2 \mu ^2-2 \mu +5 n^2+1\right)}{360 \pi ^2 (n-1) n^5}\ .
\ee
For a Weyl fermion,
\be\label{extraf}
\Delta S^f(\mu) = \frac{V_5\, \mu ^2 \left(16 \mu ^4-20 \mu ^2+135 n^4-50 \left(2 \mu ^2-1\right) n^2+7\right)}{1440 \pi ^2 (n-1) n^5}\ .
\ee
Notice that $\Delta S^f(\mu)$ is an even function of $\mu$, but $\Delta S^s(\mu)$ is not. Also note that the effective chemical potential for a dynamical field depends on the product of the R-charge and the value of the background field.
\subsection{R-symmetry chemical potential}
Following the same way of the construction of 4$d$ supersymmetric R\'enyi entropy, we may first look at the Killing spinors on a branched six-sphere $\Bbb{S}^6_n$. By solving the Killing spinor equations one can determine the R-symmetry chemical potential for the Killing spinor, $\mu(n)$. The explicit computation is performed in Appendix \ref{KSEa}.

Here we give a simple argument leading to the final result in any $d$-dimensions
\be\label{kschemical}
\mu(n) = {n-1\over 2}\ .
\ee
The argument is based on the constraints $\mu(n)$ should satisfy. We first consider the limit $n\to \infty$. In this limit the R\'enyi entropy and the extra contribution due to the nontrivial chemical potential should be of order one, since we expect to have a nontrivial Casimir energy due to the curvature scale of $\Bbb{H}^5$. Combining with the functions (\ref{extras})(\ref{extraf}), the highest power of $n$ in $\mu(n)$ should be 1. Then we consider the limit $n\to 0$. In this limit, we expect the supersymmetry to be still preserved, therefore fermions should have periodic boundary conditions in (\ref{twistkernel}), which fixes the value $|\mu|_{n\to 0}$ to be half integers. \footnote{The R-charges of the dynamical spinors and Killing spinors have the same absolute values.} Finally $\mu(n=1)=0$ should be satisfied as discussed before. Choosing a convention that $\mu(n)$ increases as $n$ increases, we eventually fix the function $\mu(n)$ as shown in (\ref{kschemical}). This is consistent with the explicit computation in $d=2,3,4,5$~\cite{Gomis:2012wy,YN,Huang:2014gca,Huang:2014pda,Hama:2014iea}. In Appendix \ref{KSEa} we demonstrate that (\ref{kschemical}) holds for $d=6$.

\subsection{Supersymmetric R\'enyi entropy}
The R-symmetry group of 6$d$ $(2,0)$ theories is $SO(5)$, which has two $U(1)$ Cartans. To compute the supersymmetric R\'enyi entropy of the tensor multiplet we first look at the R-charges $(k_1, k_2)$ of the component fields under the two Cartans, as listed in Table \ref{2charges}, where $(\psi^1, \psi^2)$ and $(\psi^3, \psi^4)$ are two Weyl fermions, and $\Phi^1:=\phi^1+i \phi^2$ and $\Phi^2:=\phi^3+i\phi^4$ are two complex scalars.

\begin{table}[htp]
\caption{charges under two $U(1)$ Cartans}
\begin{center}
\begin{tabular}{l*{7}{c}r}
              &$\psi^1$ & $\psi^2$ & $\psi^3$ & $\psi^4$ & $B_{\mu\nu}$ & $\Phi^1$ & $\Phi^2$&$\phi^5$ \\
\hline
$k_1$        & $+\frac12$ & $-\frac12$   & $-\frac12$  & $+\frac12$  &0&+1&0&0  \\
$k_2$        &  $-\frac12$ & $+\frac12$   & $-\frac12$  & $+\frac12$ &0&0&+1&0 \\
\end{tabular}
\end{center}
\label{2charges}
\end{table} 

\subsubsection{A single $U(1)$}
If we only turn on a single $U(1)$ chemical potential, for instance $A^2=0$, by the constraint (\ref{kschemical}) of the Killing spinor equation, the background field should be
\be
A^1=n-1\ .
\ee
From Table \ref{2charges}, we see that there are two Weyl fermions charged $|k_1|=1/2$ and one complex scalar charged $k_1=1$. The supersymmetric R\'enyi entropy is then computed by
\be
S_1 = S^{(2,0)_n} + \Delta S^s(\mu=n-1) + 2\Delta S^f(\mu=(n-1)/2) = {V_5\over \pi^2}\left(6n+1\over 12n\right)\ .
\ee

\subsubsection{Two same $U(1)$'s}
Now we turn on two $U(1)$ chemical potentials with the same value, $A^1=A^2$. Since the Killing spinors are charged under both two Cartans (we only consider the Killing spinor with R-charges of the same sign $|k_1+k_2|=1$), under the constraint (\ref{kschemical}) the background field should be
\be
A^1=A^2={n-1\over 2}\ .
\ee
From Tabel \ref{2charges}, we see that there are one Weyl fermion charged $|k_1+k_2|=1$ and two complex scalars charged $+1$, so the supersymmetric R\'enyi entropy is
\be\label{eq:TwoSameU1}
S_2 = S^{(2,0)}_n + 2\Delta S^s(\mu=(n-1)/2) + \Delta S^f(\mu=(n-1)/2) = \frac{\left(91 n^3+19 n^2+n+1\right) V_5}{192 \pi ^2 n^3}\ .
\ee

\subsubsection{Two generic $U(1)$'s}
We can also consider two $U(1)$ chemical potentials given by
\be
  A^1 = (n-1) a\, ,\quad A^2 = (n-1) (1-a)\, ,
\ee
where $a$ is a real deformation parameter. In this case, there are one complex scalar with chemical potential $A^1$, one complex scalar with chemical potential $A^2$ and two Weyl fermions with chemical potential $(A^1 - A^2)/2$ and $(A^1 + A^2)/2$ respectively. The supersymmetric R\'enyi entropy is
\begin{align}
  S_2 & = S^{(2,0)}_n + \Delta S^s(\mu=A^1) + \Delta S^s(\mu=A^2) + \Delta S^f(\mu=\frac{A^1 - A^2}{2}) + \Delta S^f(\mu=\frac{A^1 + A^2}{2}) \nonumber\\
  {} & = \frac{\left( C_3 n^3 + C_2 n^2 + C_1 n + C_0 \right) V_5}{12 \pi^2 n^3} \, ,
\end{align}
where the coefficients are
\begin{align}
  C_0 & = a^2 - 2 a^3 + a^4\, ,\nonumber\\
  C_1 & = a - 4 a^2 + 6 a^3 - 3 a^4\, ,\nonumber\\
  C_2 & = 1 + 3 a^2 - 6 a^3 + 3 a^4\, ,\nonumber\\
  C_3 & = 6 - a + 2 a^3 - a^4\, .
\end{align}
For $a = \frac{1}{2}$, the result is the same as Eq.~\eqref{eq:TwoSameU1} for the case with two same $U(1)$'s.

\section{Discussion}
In this note we have discussed the R\'enyi entropy and the supersymmetric R\'enyi entropy for the six-dimensional $(2, 0)$ tensor multiplet with a spherical entangling surface.  The results are consistent with the existing results in the literature. 

It would be interesting to go further to compute the supersymmetric R\'enyi entropy for the interacting $(2,0)$ theory and finally establish the TBH$_7$/qSCFT$_6$ correspondence following the same spirit of TBH$_4$/qSCFT$_3$~\cite{Huang:2014gca} and TBH$_5$/qSCFT$_4$~\cite{Huang:2014pda}.

As a by-product, we have calculated the R\'enyi entropy for a 2-form field in six dimensions. This approach could be generalized to other cases with gauge symmetry, for instance higher forms or graviton. This opens up the possibility of computing the (supersymmetric) R\'enyi entropy for more field theories.

\section*{Acknowledgement}
We are grateful for useful discussions with Andreas Gustavsson, Igor Klebanov, Shailesh Lal, Hong Liu, Robert Myers, Vasily Pestun, Soo Jong Rey, Cobi Sonnenschein and Xi Yin. YZ would like to thank Princeton University and Harvard University for their hospitality. JN was supported in part by the National Science Foundation under Grant No. PHY13-16617. YZ was supported by ``The PBC program for fellowships for outstanding post-doctoral researcher from China and India of the Israel council of higher education'' and he was also supported in part by the Israel Science Foundation (grant 1989/14 ), the US-Israel bi-national fund (BSF) grant 2012383 and the German Israel bi-national fund GIF grant number I-244-303.7-2013.

\appendix

\section{Killing spinors on $\mathbb{S}_n^6$}
\label{KSEa}
As discussed in the introduction, in order to compute the supersymmetric R\'enyi entropy, we need to solve the Killing spinor equation on the branched sphere. Let us consider the branched six-sphere $\mathbb{S}_n^6$ given by the metric
\be
  \frac{ds_6^2}{\ell^2} = d\psi^2 + \textrm{sin}^2 \psi \left[d\chi^2 + \textrm{sin}^2 \chi \left(d\rho^2 + \textrm{sin}^2 \rho \left(d\theta^2 + n^2 \textrm{sin}^2 \theta \, d\tau^2 + \textrm{cos}^2 \theta \, d\phi^2\right) \right) \right]\, ,
\ee
where $n$ is the branching parameter. The vielbeins are chosen to be
\begin{align}
  e^1 & = \ell \, d\psi\, , & e^4 & = \ell \, \textrm{sin}\psi\, \textrm{sin}\chi\, \textrm{sin}\rho\, d\theta\, ,\nonumber\\
  e^2 & = \ell \, \textrm{sin}\psi\, d\chi\, , & e^5 & = n \ell \, \textrm{sin}\psi\, \textrm{sin}\chi\, \textrm{sin}\rho\, \textrm{sin}\theta\, d\tau\, ,\\
  e^3 & = \ell \, \textrm{sin}\psi\, \textrm{sin}\chi\, d\rho\, , & e^6 & = \ell\, \textrm{sin}\psi\, \textrm{sin}\chi\, \textrm{sin}\rho\, \textrm{cos}\theta\, d\phi\, .\nonumber
\end{align}
The nonvanishing components of the spin connections are
\begin{align}
  \omega^{21}_\chi & = -\omega^{12}_\chi = \textrm{cos}\psi\, , & \omega^{41}_\theta & = - \omega^{14}_\theta = \textrm{cos}\psi\, \textrm{sin}\chi\, \textrm{sin}\rho\, ,\nonumber\\
  \omega^{31}_\rho & = -\omega^{13}_\rho = \textrm{cos}\psi\, \textrm{sin} \chi\, , & \omega^{42}_\theta & = - \omega^{24}_\theta = \textrm{cos}\chi\, \textrm{sin}\rho\, ,\nonumber\\
  \omega^{32}_\rho & = -\omega^{23}_\rho = \textrm{cos} \chi\, , & \omega^{43}_\theta & = - \omega^{34}_\theta = \textrm{cos}\rho\, ,\nonumber\\
  \omega^{51}_\tau & = -\omega^{15}_\tau = n\, \textrm{cos}\psi\, \textrm{sin}\chi\, \textrm{sin}\rho\, \textrm{sin}\theta\, , & \omega^{61}_\phi & = -\omega^{16}_\phi = \textrm{cos}\psi\, \textrm{sin}\chi\, \textrm{sin}\rho\, \textrm{cos}\theta\, ,\\
  \omega^{52}_\tau & = -\omega^{25}_\tau = n\, \textrm{cos}\chi\, \textrm{sin}\rho\, \textrm{sin}\theta\, , & \omega^{62}_\phi & = -\omega^{26}_\phi = \textrm{cos}\chi\, \textrm{sin}\rho\, \textrm{cos}\theta\, , \nonumber\\
  \omega^{53}_\tau & = -\omega^{35}_\tau = n\, \textrm{cos}\rho\, \textrm{sin}\theta\, , & \omega^{63}_\phi & = -\omega^{36}_\phi = \textrm{cos}\rho\, \textrm{cos}\theta\, , \nonumber\\
  \omega^{54}_\tau & = -\omega^{45}_\tau = n\, \textrm{cos}\theta\, , & \omega^{64}_\phi & = -\omega^{46}_\phi = - \textrm{sin}\theta\, . \nonumber
\end{align}
The Killing spinor equation for the round $\mathbb{S}^6$
\be
  \nabla_\mu \zeta := \partial_\mu \zeta + \frac{1}{4} \omega^{ab}_\mu \gamma_{ab} \zeta = -\frac{i}{2\ell} \gamma_\mu \zeta
\ee
has the solution
\be
  \zeta = e^{-\frac{i}{2} \gamma_1 \psi}\, e^{\frac{1}{2} \gamma_{12} \chi}\, e^{\frac{1}{2} \gamma_{23} \rho}\, e^{\frac{1}{2} \gamma_{34} \theta}\, e^{\frac{1}{2} \gamma_{45} \tau}\, e^{\frac{1}{2} \gamma_{36} \phi} \, \zeta_0\, ,
\ee
where $\zeta_0$ is a constant spinor, and $\gamma_{ij} := \frac{1}{2} (\gamma_i \gamma_j - \gamma_j \gamma_i)$. For the branched sphere $\mathbb{S}_n^6$, the Killing spinor equation becomes
\be\label{eq:KillingSn6}
  \partial_\mu \zeta + \frac{1}{4 n} \omega^{ab}_\mu \gamma_{ab} \zeta = - \frac{i}{2 n \ell} \gamma_\mu \zeta\, .
\ee
In a special choice of the basis
\begin{align}
  \gamma_1 & = \sigma_1 \otimes \mathbb{I} \otimes \mathbb{I}\, ,\nonumber\\
  \gamma_2 & = \sigma_2 \otimes \mathbb{I} \otimes \mathbb{I}\, ,\nonumber\\
  \gamma_3 & = \sigma_3 \otimes \sigma_1 \otimes \mathbb{I}\, ,\nonumber\\
  \gamma_4 & = \sigma_3 \otimes \sigma_2 \otimes \mathbb{I}\, ,\\
  \gamma_5 & = \sigma_3 \otimes \sigma_3 \otimes \sigma_1\, ,\nonumber\\
  \gamma_6 & = \sigma_3 \otimes \sigma_3 \otimes \sigma_2\, ,\nonumber
\end{align}
where $\sigma_i\,\, (i = 1, 2, 3)$ are the Pauli matrices, and the constant spinor
\be
  \zeta_0 = \left(\begin{array}{c}
                   0 \\ c_1 \\ c_1 \\ 0 \\ 0 \\ c_2 \\ c_2 \\ 0
                  \end{array}\right)\qquad (c_1, c_2: \textrm{constants})\, ,
\ee
Eq.~\eqref{eq:KillingSn6} is equivalent to the following Killing spinor equation, if we turn on a background gauge field $A_\mu$ and require that the solution $\zeta$ for the round $\mathbb{S}^6$ remains as a solution on the branched sphere $\mathbb{S}_n^6$:
\be
  D_\mu \zeta := \partial_\mu \zeta + \frac{1}{4} \omega^{ab}_\mu \gamma_{ab} \zeta + i A_\mu \zeta = -\frac{i}{2\ell} \gamma_\mu \zeta\, ,
\ee
where the background gauge field is
\be
  A = \frac{n-1}{2} d\tau\, .
\ee
If we choose a different Killing spinor with the same $\gamma$-matrices and the same constant spinor $\zeta_0$:
\be
  \widetilde{\zeta} = e^{\frac{i}{2} \gamma_1 \psi}\, e^{\frac{1}{2} \gamma_{12} \chi}\, e^{\frac{1}{2} \gamma_{23} \rho}\, e^{\frac{1}{2} \gamma_{34} \theta}\, e^{\frac{1}{2} \gamma_{45} \tau}\, e^{\frac{1}{2} \gamma_{36} \phi} \, \zeta_0\, ,
\ee
it satisfies another Killing spinor equation
\be
  D_\mu \widetilde{\zeta} := \partial_\mu \widetilde{\zeta} + \frac{1}{4} \omega^{ab}_\mu \gamma_{ab} \widetilde{\zeta} + i A_\mu \widetilde{\zeta} = \frac{i}{2\ell} \gamma_\mu \widetilde{\zeta}
\ee
with the same background gauge field $A_\mu$ as before.

\bibliographystyle{utphys}
\bibliography{RenyiPaper}

\end{document}